# SCIENCE PARAMETRICS FOR MISSIONS TO SEARCH FOR EARTH-LIKE EXOPLANETS BY DIRECT IMAGING

Robert A. Brown

Space Telescope Science Institute, 3700 San Martin Drive, Baltimore, MD 21218

rbrown@stsci.edu

ABSTRACT

We use $N_t$, the number of exoplanets observed in time $t$, as a science metric to study direct-search missions like *Terrestrial Planet Finder*. In our model, $N$ has 27 parameters, divided into three categories: 2 astronomical, 7 instrumental, and 18 science-operational. For various "27-vectors" of those parameters chosen to explore parameter space, we compute design reference missions to estimate $N_t$. Our treatment includes the recovery of completeness $c$ after a search observation, for revisits, solar and antisolar avoidance, observational overhead, and follow-on spectroscopy. Our baseline 27-vector has aperture $D = 16$m, inner working angle $IWA = 0.039"$, mission time $t = 0$–5 years, occurrence probability for earthlike exoplanets $\eta = 0.2$, and typical values for the remaining 23 parameters. For the baseline case, a typical five-year design reference mission has an input catalog of ~4700 stars with nonzero completeness, ~1300 unique stars observed in ~2600 observations, of which ~1300 are revisits, and it produces $N_1$ ~50 exoplanets after one year and $N_5$ ~130 after five years. We explore offsets from the baseline for ten parameters. We find that $N$ depends strongly on $IWA$ and only weakly on $D$. It also depends only weakly on zodiacal light for $Z < 50$ zodis, end-to-end efficiency for $h > 0.2$, and scattered starlight for $\zeta < 10^{-10}$. We find that observational overheads, completeness recovery and revisits, solar and antisolar avoidance, and follow-on spectroscopy are all important factors in estimating $N$.

*Subject headings*: instrumentation: high angular resolution, methods: statistical, planetary systems, planets and satellites: detection

# 1. INTRODUCTION

Teams of astronomers and optical experts are now designing future space telescopes and instruments for high-dynamic-range imaging to search for extrasolar planets. The goal is to find and characterize earthlike planets, with the size, atmospheric composition, and temperature of earth—possible habitats of life. *Terrestrial Planet Finder* is the prototype of such missions: aperture in the range $D = 4$–$64$ m; a nominal operating wavelength $\lambda = 760$ nm, which allows diagnostic spectroscopy of methane and oxygen; inner working angle $IWA = 0.02$–$0.31''$, which is the effective radius of the central field obscuration; and starlight suppression $\zeta = 10^{-7}$–$10^{-11}$ by an internal coronagraph, where $\zeta$ is the intensity of scattered starlight expressed in units of the theoretical central intensity of the stellar Airy disk.

Because of the great expense and technical challenge of mounting a mission like *Terrestrial Planet Finder*, competing concepts must be tested with a science metric that probes the many optical and operational issues. The most compelling science metric for this purpose is the estimated number of planets $N_t$ that would be discovered in mission time $t$.

The goals of metrical analysis are at least threefold: to better understand the basic character of such a complex mission with many moving parts, to inform tradeoffs and reductions of scope in a cost-sensitive environment, and to set expectations that could reasonably be fulfilled if a mission is implemented.

This paper reports a parametric analysis using the formalism of the design reference mission, which is an optimized list of observations that simulate the mission as a whole, including random discoveries. In particular, we use design reference missions to estimate N as a function of 27 mission parameters, which are divided into three categories: 2 are astronomical, 7 are instrumental, and 18 are science-operational. (See Table 1.) We use the term "27-vector" to refer to a complete set of values for this set of 27 mission parameters.

Our model treats a variety of science-operational aspects, including (1) waiting for available completeness $c$ to recover after a limiting search observation that did not detect a planet (Brown & Soummer 2010); (2) solar and antisolar pointing avoidance; (3) overhead time $t_{o/h}$ per observation, and (4) "filler" days that could be used for other observing programs when no candidate target star is searchable—a situation that can occur for a short input catalog when all candidate targets are either forbidden by the pointing restrictions or are still waiting for completeness $c$ to recover.

A limiting search observation is an exposure to reach the systematic limit of detection, $\Delta mag_0$, such that detecting a fainter source is assumed not to be possible. $\Delta mag$ is the flux ratio between star and planet expressed in stellar magnitudes.

Completeness $c$ is defined as the fraction of planets of interest that would be detected according to two criteria: $s > IWA$, which means the planet is not obscured (Brown 2004a), and $\Delta mag < \Delta mag_0$, which means that the planet is brighter than the noise floor (Brown 2005). The apparent separation $s$ is the angle between the planet and the host star as seen by the observer.

The purpose of solar and antisolar pointing restrictions is to avoid sunlight scattering into the optical path of the target star during an exposure. In the general case, pointing might be restricted in both the solar and antisolar directions.

Our baseline 27-vector is a mission with $D = 16$m, $IWA = 0.039"$, $t = 0$–5 years,

occurrence probability η = 0.2, and typical values for the other 23 parameters. A design reference mission for the baseline case involves ~4700 candidate stars with nonzero completeness, ~1300 unique stars observed in 2600 observations, of which 1300 are revisits, and produces an estimated $N_1$ ~50 exoplanets after one year and $N_5$ = 130 after five years of elapsed mission time. These statistics vary for parameters away from the baseline.

To investigate the parametric variations of the science metric $N$, we compute design reference missions for a suite of diagnostic 27-vectors in the vicinity of the baseline. In §2, we define and describe the components of our computations—a *mise en place*. In §3, we give the step-by-step recipe for computing a design reference mission and estimating $N$. In §4, we summarize our findings, and in §5 we offer conclusions.

## 2. MISE EN PLACE

In days of yore, every observatory dome had a logbook in which the observer recorded, usually in ink, each observation with the telescope—object name, celestial coordinates, instrument used, exposure time, universal time, and comments, which were often about the weather. Each entry in the logbook also had an unrecorded backstory: why a target was selected and what was already known about it. A follow-on narrative was also implied but unrecorded: the results after data reduction, calibration, and analysis. To better understand the research reported here, about a great space telescope of the future, it may be helpful from time to time to ponder the analogy between an old-time logbook and a design reference mission. We could start by noting that each of the 27 parameters in Table 1 has a valid counterpart in the classical setting. Other correspondences will come to mind, such as between the old-time observer in the telescope dome and the scheduling algorithm.

## 2.1 ENGINE FOR DESIGN REFERENCE MISSIONS

Our engine for producing design reference missions is actually MATHEMATICA computer code. The input to the code is a 27–vector, and the output is a design reference mission, including results for $N_t$. The design reference mission is literally a table—here, table with 12 columns—with as many rows as the number of observations it describes. Each row comprises 12 data items: (1) *Hipparcos* number HIP; (2) mission day $t$ on which the observation is performed; (3) the number of visits to this target up to now; (4) the value of $c$ for this observation (Brown 2005); (5) the total completeness $C$ harvested by all searches of this target up until now; (6) the Bayesian probability

$$\mathcal{P} \equiv \frac{\eta c}{1 - \eta C} \qquad (1)$$

that this observation will discover a planet (Brown & Soummer 2010); (7) whether a planet is in fact discovered, true or false, as determined by a draw from a Bernoulli random deviate of probability $\mathcal{P}$; (8) total discoveries by all observations until now; (9) the merit function for this observation

$$\mathcal{Z} \equiv \frac{\mathcal{P}}{\tau_{\mathrm{LSO}} + t_{\mathrm{o/h}}} \; ; \qquad (2)$$

(10) exposure time $t_{\mathrm{LSO}}$ of a limiting search observation (see § 2.10); (11) the actual exposure time $t_{\mathrm{spc}}$, including spectroscopy if a planet has been discovered; and (12) the total time cost of the observation, including observational overhead (see § 2.7).

h is the assumed occurrence rate of the planets of interest. $\mathcal{Z}$ might be called the information rate, discovery rate, or benefit-to-cost ratio of the observation.

Table 1 is a list of the parameters in a 27-vector and the values they take on in the baseline.

Table 2 is a truncated example of a design reference mission for the baseline, giving the first ten and last ten lines.

Table 3 summarizes the results of design reference missions for the baseline and diagnostic offsets of the primary parameters, *D* and *IWA*. To build Table 3, we compute 240 design reference mission *ab initio* for each 27-vector and averaged the results.

Tables 4–11 summarize the results for diagnostic offsets for eight secondary parameters.

## 2.2 SCHEDULING ALGORITHM

In the scenario of a design reference mission, the scheduling algorithm plays the role of the observer, continually deciding the question of what star to search next. The result is always the qualified star offering the highest value of the merit function $\mathcal{Z}$. A "qualified star" is both ready and able—"ready" if the completeness has recovered from any prior observation and "able" if the rules for solar and antisolar avoidance permit the telescope to point at the target.

If a promising feature is discovered during a limiting search observation, the scheduling algorithm extends the current observation, converting it to a longer spectroscopic exposure.

## 2.3 TWO TYPES OF OBSERVATIONS

Each new observation in a design reference mission starts as a limiting search observation. That is, the scheduling algorithm initially assumes the presence of a planetary companion somewhere in the detection zone of the instrument. For planning purposes, the scheduling algorithm assumes the worse-case scenario, that the planet is a limiting source, at the noise floor, and $\Delta mag_0$ magnitudes fainter than the host star. The plan—if no planet appears on the detector—is to conduct an exposure of length $t_{LSO}$, which, according to the exposure time calculator, will achieve $S/N_{img}$ through a spectral bandpass of $\lambda/\mathcal{R}_{img}$, for a source of magnitude $mag_s + \Delta mag_0$. The exposure time calculator is defined in §2.7.

If no planet is detected after $t_{LSO}$ has elapsed, the scheduling algorithm books the total time cost of the observation as $\tau_{LSO} + t_{o/h}$, and moves on to the next star.

If a planet *is* detected, then the scheduling algorithm draws a detectable random planet from the appropriate distribution of orbital elements, defined in §2.5, and declares that *this* planet is the discovered planet. We now recognize that a design reference mission is a random variable.

Next, the scheduling algorithm computes the spectroscopic exposure time $\tau_{spc}$ needed to achieve signal-to-noise ratio $SNR_{spc}$ through a bandpass of $\lambda/\mathcal{R}_{spc}$ for this particular planet and host star. The scheduling algorithm books the time cost of the observation as $\tau_{spc} + t_{o/h}$, tallies the discovery, and moves on to the next star.

In the case of no discovery, the searched star remains in play, but it is not yet ready to observe again until $c$ has recovered. If a discovery *is* made, the scheduling algorithm takes the star out of play.

The good reason for immediately following up a possible discovery with a spectroscopic observation is that the opportunity is perishable. Planets move, and the scheduling algorithm has no information about where the source will be in the future. The meager orbital information available at the time of discovery is the one measurement of $s$, which is insufficient to predict the planet's future position. Indeed, research is still needed to address the variety of post-detection issues, including source characterization, establishing companionship, and estimating orbital elements (Brown, Shaklan, and Hunyadi 2006).

Clearly, obtaining the spectrum has great scientific importance, and it will be hard to obtain if not taken immediately upon discovery.

## 2.4 COMPLETENESS

Through the detection probability $\mathcal{P}$ and the merit function $\mathcal{Z}$ in Equations (1–2), the scheduling algorithm relies on data products related to completeness. We estimate $c$ by Monte Carlo experiments involving 10, 000 random planets, which revolve around the star and change in brightness according to their orbital elements and the current time. Each successive search of a star harvests a different value of $c$ equal to the number of planets not ruled out by previous searches—but detectable by this one—divided by 10, 000.

Brown and Soummer (2010) develop the concept of dynamic completeness, which recognizes that the positions of all planets that were not ruled out by previous searches will continue to evolve according to their orbital elements, and possibly become observable by a future search.

$c$ is zero immediately after an unproductive search, and so the scheduling algorithm must plan a minimum delay for the next search of this star, to allow $c$ to recover. For habitable-zone orbits defined by equilibrium temperature, the recovery time depends on the mass and luminosity of the host star. In the current research, we pivot from Brown and Soummers results for HIP 29271, in their Figure 1, which suggests a nominal recovery time $t_{rcv} = 58$ days for that star. Our catalog values of mass and luminosity for HIP 29271 are 0.91 $M_\odot$ and 0.83 $L_\odot$. Therefore, our scaled estimate of the recovery time for another star, of mass $M$ and luminosity $L$, is:

$$t_{rcv} = 58 \left(\frac{L}{0.83}\right)^{3/4} \left(\frac{M}{0.91}\right)^{1/2} \text{ days} . \qquad (3)$$

One task of the scheduling algorithm is to start the recovery clock running for any star with an unproductive search, and to bring that star back into play for selection when $t_{rcv}$ has elapsed.

## 2.5 PLANETS

The planets are represented by a distribution of planetary orbits and the assumed values of two photometric quantities: planetary radius $R = 1$ earth radius and geometric albedo $p = 0.3$. We assume a Lambertian phase function.

For the completeness calculations, we prepare random samples of planets by drawing values of six orbital elements from random deviates with the same parameters used in Brown (2005) and Brown and Soummer (2010). The semimajor axis $a$ is in the habitable zone, uniformly distributed in the range $0.7\sqrt{L} \leq a \leq 1.5\sqrt{L}$. The eccentricity $e$ is uniformly distributed in the range $0 \leq e \leq 0.35$. The initial mean anomaly ($M_0$) is uniform in the range $0 \leq M_0 \leq 2\pi$. The three Euler angles to orient the orbit in space—inclination $i$, argument of periapsis $\omega_0$, and position angle of the ascending node $\Omega$—are distributed uniformly on the sphere.

For completeness calculations, we create 10,000 random planets for each 27-vector and each star, and then compute each planet's position and brightness—particularly $s$ and $\Delta mag$—at any future time. For the orbital calculations, we use the computational recipe described in Brown (2004b), §3.1.

## 2.6 INPUT CATALOGUE OF STARS

We use an input catalog of 18,865 stars assigned values of $L$ and $M$, distance $d <$ 100 pc, visual magnitude $V$, color $0.3 \leq B - V \leq 2$, and luminosity class in the range IV–V. Our sample was drawn from the NASA Star and Exoplanet Database (NStED), which has since vanished.

We assigned $I$ magnitudes to each star using a fifth-order fit to the empirical relationship between $V$ and $I$ in Neill Reid's photometry at

http://www.stsci.edu/~inr/phot/allphotpi.sing.2mass:

$$I = V - (1.58344 - 9.35193 (B - V)$$
$$+ 26.22379 (B - V)^2 - 30.73087 (B - V)^3 \quad (4)$$
$$+ 16.51837 (B - V)^4 - 3.20157 (B - V)^5)$$

We remove from consideration all stars with habitable zones permanently obscured by the central field stop. The maximum possible apparent separation of a planet from its host star is

$$s_{max} = ((a_{max} = 1.5 \text{ AU}) (1 + (e_{max} = 0.35))) \sqrt{L} / d. \quad (5)$$

Therefore, at the outset we can eliminate as invalid all stars for which $s_{max} < IWA$.

## 2.7 OVERHEAD TIME

$t_{o/h}$ includes all time costs of a search observation other than the exposure time $\tau$. Typical overhead tasks are repointing the telescope, fine alignment of the optics, equilibrating the system when heat loads change, and any unique calibrations that must be charged to this observing program. The total time cost of an observation is $\tau + t_{o/h}$. We adopt the baseline value $t_{o/h} = 0.5$ days, which is typical of current NASA studies, and we explore the range $0 \le t_{o/h} \le 2$ days using parametric offsets (see Table 8 and Figure 8).

## 2.8 SOLAR AND ANTISOLAR AVOIDANCE

As shown in Table 1, we have adopted the values $\gamma_1 = 45°$ for solar avoidance and $\gamma_2 = 180°$ for antisolar avoidance (*i.e*, no antisolar restriction). These choices are typical of current NASA studies of similar missions.

To compute the pointing restrictions, we use a right-handed, rectangular, ecliptic coordinate system, centered on the observer, with the sun fixed. The unit vector in the direction of the sun is always $\boldsymbol{u}_{\text{sun}} = (-1, 0, 0)$. We convert the right ascension and declination of the target star into ecliptic longitude $\phi$ and latitude $\theta$, then we convert the unit vector to the target into rectangular, ecliptic coordinates:

$$\boldsymbol{u}_{\text{target}} = \begin{pmatrix} \cos\theta \cos\left(\phi - 2\pi \frac{t + t_{\text{start}} - t_{\text{ve}}}{365.2 \text{ days}} - \pi\right), \\ \cos\theta \sin\left(\phi - 2\pi \frac{t + t_{\text{start}} - t_{\text{ve}}}{365.2 \text{ days}} - \pi\right), \\ \sin\theta \end{pmatrix}, \qquad (6)$$

where $t$ is the mission time in days, $t_{\text{start}}$ is the Julian day of mission start, and $t_{\text{ve}}$ is the Julian day of a vernal equinox.

The dot product of the two unit vectors is $\cos\alpha = \boldsymbol{u}_{\text{sun}} \cdot \boldsymbol{u}_{\text{target}}$, where $\alpha$ is the angle between the sun and the target as seen by the observer. The scheduling algorithm tests pointing the restriction $\gamma_1 < \alpha < \gamma_2$ for candidate targets at the current mission time $t$.

## 2.9 NOISE FLOOR

The noise floor $\Delta mag_0$ is the systematic limit of detectability for the faintest sources. The limit is due to the temporal instability of the optical system (Brown 2005). The picture is that $S/N$ increases as $\tau^{1/2}$ for sources with $\Delta mag < \Delta mag_0$, but that the detection of any fainter source, $\Delta mag \geq \Delta mag_0$, is impossible with any amount of exposure time.

## 2.10 EXPOSURE TIME CALCULATION

For a host star of magnitude $mag_s$, the exposure time $\tau$ for a planet of magnitude

$mag_s + \Delta mag$ is the time needed to achieve the desired signal-to-noise ratio $S/N$:

$$S/N = \frac{S}{\sqrt{S + q\mathcal{N}}}, \tag{7}$$

where the signal S is the total of planetary counts, and the noise N is the total of the non-planetary counts. Both counts are gathered in the virtual photometric aperture, which comprises $1/\Psi$ pixels, where $\Psi$ is the sharpness (Burrows 2003; Burrows, Brown, and Sabatke 2006). The factor $q$ allows for possible roll convolution, where a number $q$ of statistically equivalent images of pure background are arithmetically combined—i.e. differenced, shifted, and added—possibly having been obtained at different roll angles of the telescope to disambiguate speckles. In our current research, $q = 1$.

The signal counts are

$$S = hF_p \frac{\lambda}{\mathcal{R}} \frac{\pi D^2}{4} \tau, \tag{8}$$

where the planetary flux is

$$F_p = F_0 \, 10^{-\frac{mag_s + \Delta mag}{2.5}}, \tag{9}$$

$h$ is the end-to-end efficiency, $\lambda$ is the operational wavelength, $\mathcal{R}$ is the spectral resolving power, $D$ is the diameter of the aperture, and the zero point is $F_0 = 4885$ photons cm$^{-2}$ nm$^{-1}$ sec$^{-1}$ for $\lambda = 760$ nm.

N is the sum of contributions from zodiacal light $Z$, scattered starlight $\zeta$, dark noise $\upsilon$, and read noise $\sigma$:

$$\mathcal{N} = \mathcal{N}_Z + \mathcal{N}_\zeta + \mathcal{N}_\upsilon + \mathcal{N}_\sigma. \tag{10}$$

We measure $Z$, the total zodiacal light—local plus extrasolar—in units of "zodis," the surface brightness of the local zodiacal light, $mag_Z = 23$ mag arcsec$^{-2}$ (Leinert et al.

1998). The counts from zodiacal light are

$$\mathcal{N}_Z = Z h F_0 \frac{10^{\frac{mag_Z}{2.5}}}{(4.848 \times 10^{-6})^2} \frac{1}{\Psi} \frac{\lambda^2}{4D^2} \frac{\lambda}{\mathcal{R}} \frac{\pi D^2}{4} \tau$$
$$= \frac{5.271 \, Z h F_0 \lambda^3}{\Psi \mathcal{R}} \tau.$$
(11)

The counts from scattered starlight are

$$\mathcal{N}_\zeta = \zeta h F_0 10^{\frac{mag_s}{2.5}} \frac{\pi D^2}{4\lambda^2} \frac{1}{\Psi} \frac{\lambda^2}{4D^2} \frac{\lambda}{\mathcal{R}} \frac{\pi D^2}{4} \tau$$
$$= \frac{0.1542 \, \zeta h F_0 10^{\frac{mag_s}{2.5}} \lambda D^2}{\Psi \mathcal{R}}.$$
(12)

The dark and read-noise counts are

$$\mathcal{N}_\upsilon = \frac{\upsilon}{\Psi} \tau \tag{13}$$

and

$$\mathcal{N}_\sigma = \frac{\sigma^2}{\Psi t_r} \tau, \tag{14}$$

where $t_r$ is the cadence of reading out the detector.

For the task of estimating the exposure time to achieve $S/N$, we can substitute it for $S/N$ and solve the equation for $\tau$, obtaining

$$\tau = (S/N)^2 \, 1.121 \times 10^{-16+0.4(mag_s + \Delta mag - mag_Z)} \frac{\mathcal{R}}{h \Psi D^2 F_0} \times$$

$$\times \begin{cases} 1.135 \times 10^{16+0.4(mag_Z)} \frac{\Psi}{\lambda} \\ + 1.446 \times 10^{16+0.40(mag_s + \Delta mag + mag_Z)} \frac{2\mathcal{R}}{h\lambda^2 D^2 F_0} \left( \upsilon + \frac{\sigma^2}{t_r} \right) \\ + 2.229 \times 10^{15+0.4(\Delta mag + mag_Z)} \frac{2\zeta}{\lambda} \\ + 1.208 \times 10^{12+0.4(\Delta mag + mag_s)} \frac{2\lambda Z}{D^2} \end{cases}. \quad (15)$$

## 2.11 DETECTOR

In our treatment, the parameters unique to the detector are listed in rows 5, 7, 8, and 13 in Table 1: sharpness $\Psi = 0.08$, dark noise $\upsilon = 0.00055$ counts/sec, read noise $\sigma = 2.8$ $\sqrt{\text{counts}}$/read, and readout cadence $t_r = 2000$ sec. These are typical values, adopted in current NASA studies. We assume pixels that critically sample the point-spread function at the operating wavelength $\lambda = 760$ nm, which means the angular subtense of the pixel width is $\lambda/(2D)$.

## 3. COMPUTING DESIGN REFERENCE MISSIONS

In this section, we describe the procedure for computing a design reference mission for a given 27-vector of parameters. We use the baseline defined in Table 1 as an example.

Step 1. Restrict attention to the stars with resolved habitable zones according to $s_{max} < IWA$ and Equation (5). For the baseline, some 9,108 stars out of 18,865 stars in the input catalog pass this test.

Step 2. Compute an exhaustive completeness vector for each star, as follows. Generate 10,000 random planets. Test each one to find whether or not it satisfies the two criteria for detection at the current time: brightness above the noise floor ($\Delta mag < \Delta mag_0$) and position unobscured ($s > IWA$). The virgin completeness is the total number of planets satisfying these criteria divided by 10,000. Next, remove the currently detectable planets from consideration and advance the clock by a large value, say $10^9$ days. Such a long delay allows $c$ to fully recover. Recompute $c$; the result is the second element in the completeness vector. Repeat these steps until all the planets ever detectable are exhausted, which is signified by a returned value of $c = 0$. The $i^{th}$ entry in the completeness vector of a star is the value of $c$ that the scheduling algorithm will use in Equation (1) when it computes the merit function $\mathcal{Z}$ for what would be—if it is selected—the $i^{th}$ search of the star. The value of $C$ to use in Equation (1) is the sum total of the first $i - 1$ entries in the completeness vector.

The completeness vector is a random variable, which is one reason a design reference mission is a random variable. Other randomness is introduced by the scheduling algorithm decision whether or not a planet has been detected (see Step 8 below).

Step 3. Remove all stars with a completeness vector that is identically zero, which can happen even when the habitable zone is resolved if the probability is less than $10^{-4}$ that a random planet will satisfy both detection criteria—adequate separation and brightness. This is a large effect. In a typical run with baseline parameters, only 4721 stars of the original 9108 stars remain in play after those with zero completeness are removed.

Step 4. For each star, compute the completeness recovery time $t_\text{rc}$ and the exposure time $\tau_\text{LSO}$ from Equation (3) and Equation (15), respectively.

Step 5. Sort the stars in descending order of $\mathcal{Z}$ as computed from Equations (1–2) for the first search of each star. This sorting produces a typical stack of 4721 stars, which is the starting point for building the design reference mission line by line. On top of the stack is the star with the highest merit.

Step 6. The scheduling algorithm tests the star on top of the stack for two necessary conditions: the pointing must be permitted at the current mission time $t$ and the recovery time $t_\text{rc}$ from any prior search must have elapsed. If these two conditions are satisfied, the star at the top of the stack is selected for the next search. If either condition is not satisfied, the scheduling algorithm moves down the stack and chooses the first star that does satisfy the two conditions. If the scheduling algorithm gets to the bottom of the stack and finds no selectable star, it advances the mission time by one day and tries again, starting at the top of the stack. The time skipped over would be available for filler observations to serve another observing program.

Step 7. The scheduling algorithm determines whether a discovery has been made by figuratively flipping a biased coin, with probability of "yes" equal to the value of $\mathcal{P}$ in Equation (1), for the values of $c$ and $C$ of this search.

Step 8. If "no," a discovery has *not* been made, then jump to Step 9. If "yes," a discovery *has* been made, the scheduling algorithm executes a series of substeps: (1) Draw one random planet for this star. If this planet is detectable at the current time by the criteria for $\Delta mag$ and $s$, then choose this planet to be the planet found. If the planet drawn is not detectable, then try again—draw another random planet—until a detectable one is found. (2) Compute the actual exposure time for spectroscopy of this planet according to parameters 15 and 17 in Table 2. (3) Remove this star from the stack. (4) Perform the bookkeeping, which means filling out the next row in the design reference mission. (5) Advance the mission clock by the total time cost of this observation, $\tau_{spc} + t_{o/h}$, as described in §2.3. (6) Loop back to Step 6 and repeat until the mission clock runs out.

Step 9. With no discovery, the scheduling algorithm performs a different series of substeps to reinsert the star into the stack for future consideration: (1) Pull up the next element in this star's completeness vector, after the one most recently used. Use this value of $c$—and $C$, the total of all prior values of $c$—to compute a new value of $\mathcal{Z}$ for this star. (2) Start the recovery clock running for this star. (3) Insert this star back into the stack at its new position, ranked according to its new value of $\mathcal{Z}$. (4) Advance the mission clock by the time cost of this observation, $\tau_{LSO} + t_{o/h}$. (5) Perform the bookkeeping, creating a new row in the design reference mission. (6) Loop back to Step 6 and repeat until the mission clock runs out.

Table 11 is a snapshot of a typical design reference mission for the baseline, showing the first ten and last ten lines. The meanings of the 12 columns are discussed in §2.1. Our interest centers at first on column 2 and 8: the mission time $t$ and the total number of planets observed—the science metric $N$. Figure 2 shows these curves for the baseline and other offsets of the major parameters, $D$, $IWA$, and $t$, as listed in Table 2. These results are the average of 240 design reference missions for each 27-vector. For the offsets of secondary parameters shown in Tables 4–11 and Figures 4–11, we averaged 24 design reference missions for each 27-vector.

We see in Table 3, for the baseline 27-vector (Line 6), that 1271 stars are actually observed out of the 4721 stars in the original stack. Also, 1338 of the total 2576 observations are revisits. Also not that no filler observations were needed for the baseline.

## 4. PARAMETRICS

We explore offsets from the baseline for 11 parameters: $D$, $IWA$, $t$, $\eta$, $Z$, $h$, $\zeta$, $\Delta mag_0$, $t_{o/h}$, $\mathcal{R}_{spc}$, and the desired $S/N_{spc}$ for spectroscopic characterization of any discovered planets.

The results for the major parameters, $D$, $IWA$, and $t$ are summarized in Figures 1–3 and Table 3. If we regard $D$ and $IWA$ as independent parameters, which we do, and if typical exposure times $\tau$ are shorter than $t_{o/h}$, which they are for the baseline, then $N$ depends only weakly on $D$, but strongly on $IWA$. The weak dependence on $D$ is because increasing the collecting area does not significantly reduce the total time per observation, which is dominated by $t_{o/h}$.

The two great advantages of decreasing $IWA$—whether $D$ changes or not—are accessing more stars and generally increasing search completeness.

Tables and Figures 4–11 show the parametric results for the eight secondary parameters we explore. The cases are treated as offsets from the baseline, which is why one point in the figures is always a cross, signifying the baseline case.

Table and Figure 4 shows that $N$ is strictly proportional to $\eta$, which is no surprise, but good to keep in mind. We have no control over $\eta$.

Table and Figure 5 show the results for zodiacal light $Z$. They show that $N$ depends only weakly on $Z$, showing no effect for $Z < 10$ zodis, and only a 15%, 30%, and 40% reduction in $N$ for $Z = 50$, 100, and 200 zodis, respectively. This result suggests that searching for simple infrared excesses could identify target stars with intolerable values of $Z$ (Moro-Martín 2014).

Table and Figure 6 show results for the end-to-end efficiency $h$. The time cost of an observation is $\tau + t_{o/h}$. For large $h$, $t_{o/h}$ dominates $\tau$, and in that case improving efficiency offers little benefit in terms of the science metric $N$. The situation reverses for small $h$, when $\tau$ controls the time cost per observation, and thereby controls the total number of observations and therefore the value of $N$. In this regime, $N$ depends strongly on $h$, and in the limit as $h \to 0$, $N \to 0$.

Table and Figure 7 show results for scattered starlight. Decreasing $\zeta$ from $10^{-10}$ to $10^{-11}$ does not increase $N$, but increasing it to $10^{-9}$, $10^{-8}$, or $10^{-7}$ reduces $N$ by 30%, 65%, or 100%, respectively.

Table and Figure 8 show results for observational overhead $t_{o/h}$. Reducing $t_{o/h}$ from the baseline 0.5 days to 0 days increases $N$ by 50%, while increasing $t_{o/h}$ from 0.5 to 2 days reduces $N$ by 50%. Observational overhead is important.

Table and Figure 9 show results for noise floor $\Delta mag_0$. Below $\Delta mag_0 = 26.5$, decreasing $\Delta mag_0$ reduces $N$ because the photometric completeness of all observations is reduced. Above $\Delta mag_0 = 26.5$, increasing $\Delta mag_0$ reduces $N$ because limiting searches demand more exposure time $t_{LSO}$ to reach the lower noise floor. A science requirement on $\Delta mag_0$—which this research suggests should be $\Delta mag_0 = 26.5$—is translated by the exposure time calculator into a technical requirement on temporal stability. Because temporal stability is a complex topic involving the whole observatory—and because it is a cost driver—our results for the variation of $N$ with $\Delta mag_0$ should be useful.

Table and Figure 10 show results for spectroscopic resolving power $\mathcal{R}_{spc}$. $N$ decreases linearly with $\mathcal{R}_{spc}$, but only weakly; if we change from $\mathcal{R}_{spc}$ from 17.5 to 280, $N$ decreases by 30%. $N$ declines as $\mathcal{R}_{spc}$ increases because of the time penalty to achieve the same $S/N$. The reduction in $N$ is modest because of a dilution effect: only a small fraction of all observations are affected by $\mathcal{R}_{spc}$—those producing a discovery.

Table and Figure 11 show the results for the signal-to-noise of spectroscopy $S/N_{spc}$. Above the baseline value of $S/N_{spc}$, $N_5$ is inversely proportional to $S/N_{spc}$.

## 5. SUMMARY

Metrical analysis of mission parameters using the formalism of the design reference mission offers insights into the basic character of missions like *Terrestrial Planet Finder* that search for exoplanets by direct imaging.

If search exposure times $\tau$ are short compared with observational overhead $t_{o/h}$, and if aperture $D$ is treated as independent of inner working angle $IWA$, then $D$ has a minor effect on the science metric $N$ compared with $IWA$, which has a strong effect. In this regime, the benefits of reducing $IWA$ are greater completeness and more stars with resolved habitable zones, while little benefit accrues from shortening exposure times by increasing $D$ if $t_{o/h}$ dominates $\tau$.

Other consequences of $t_{o/h}$ dominating $\tau$ include the reduced importance of end-to-end efficiency for $h > 0.2$, scattered starlight for $\zeta < 10^{-10}$, and zodiacal light for $Z < 50$ zodis.

If mission cost depends more strongly on *D* than on *IWA*, the same science may be achievable at lower cost if *D* and *IWA* can be simultaneously reduced.

The relative immunity to $Z < 50$ suggests that an advance survey of stars looking for infrared excesses could weed-out stars with potentially intolerable zodi. Observations with *Herschel* and *Spitzer* suggest that cold disks in outer planetary systems are common and may correlate with the occurrence of terrestrial planets and zodiacal dust in the inner planetary system (Moro-Martín 2014).

This research reaffirms the importance of science-operational issues in defining and optimizing missions to directly detect exoplanets (Brown, Shaklan, and Hunyadi 2006).

I thank Stuart Shaklan for many insightful comments and questions about this research. I thank Marc Postman and Sara Seager for their encouragement. I thank Margaret Turnbull for her advice on stars. I thank Amaya Moro-Martín for sharing her views on the relationships between infrared excesses, debris disks, and exozodiacal light. I thank Sharon Toolan for her help with the manuscript.

Table 1. 27 parameters of science metric $N$.

| | | parameter | symbol | baseline value | range | units |
|---|---|---|---|---|---|---|
| astronomical | 1 | occurrence probability | $\eta_{true}$ | 0.20 | 0.05–0.8 | |
| | 2 | zodi | $Z$ | 7 | .875–56 | 23 mag/arcsec$^2$ |
| instrumental | 3 | aperture | $D$ | 16 | 4–64 | meters |
| | 4 | inner working angle | $IWA$ | 0.039 | 0.02–0.31 | arcsec |
| | 5 | sharpness | $\Psi$ | | 0.08 | |
| | 6 | end-to-end efficiency | $h$ | 0.2 | 0.05–0.80 | |
| | 7 | dark noise | $\upsilon$ | | 0.00055 | counts/sec |
| | 8 | read noise | $\sigma$ | | 2.8 | $\sqrt{counts}$/read |
| | 9 | scattered starlight | $\zeta$ | $10^{-10}$ | $10^{-7}$–$10^{-11}$ | Airy peak intensity |
| science operational | 10 | operational wavelength | $\lambda$ | | 760 | nm |
| | 11 | limiting delta magnitude | $\Delta mag_0$ | 26 | 23–27.5 | |
| | 12 | observational overhead | $t_{o/h}$ | 0.5 | 0–2 | days |
| | 13 | readout cadence | $t_r$ | | 2000 | seconds |
| | 14 | resolution, imaging | $\mathcal{R}_{img}$ | | 5 | |
| | 15 | resolution, spectroscopy | $\mathcal{R}_{spc}$ | 70 | 17.5–280 | |
| | 16 | S/N, imaging | $SNR_{img}$ | | 8 | |
| | 17 | S/N, spectroscopy | $SNR_{spc}$ | 8 | 8–160 | |
| | 18 | rolls | $q$ | | 1 | |
| | 19 | mission start | $t_{start}$ | | 2460310 (1/1/2024) | Julian day |
| | 20 | mission duration | $t_{start} - t_{stop}$ | | 5 (1826) | years (days) |
| | 21 | planetary radius | $R$ | | 1 | earth radii |
| | 22 | geometric albedo | $p$ | | 0.3 | |
| | 23 | semimajor axis | $a$ | | (0.7–1.5) $\sqrt{L}$ | AU |
| | 24 | eccentricity | $e$ | | 0–0.35 | |
| | 25 | solar avoidance angle | $\gamma_1$ | | 45° | degrees |
| | 26 | antisolar avoidance angle | $\gamma_2$ | | 180° | degrees |
| | 27 | mission time | $t$ | 1826 | $0 \leq t \leq 1826$ | days |

Notes. $a$ and $e$ are assigned random values uniformly distributed in the indicated rages.

The pole of the orbit is distributed uniformly on the sphere.

Table 2. A typical five-year design reference mission for the baseline parameters (truncated).

| Obser-vation | HIP | $t$ | Visits | $c$ | $C$ | $\mathcal{P}$ | Dis-covery | Total dis-coveries | $\mathcal{Z}$ | $\tau_{LSO}$ | $\tau$ | Total time cost |
|---|---|---|---|---|---|---|---|---|---|---|---|---|
| 1 | 16537 | 0.5 | 1 | 0.875 | 0.875 | 0.175 | yes | 1 | 0.346 | 0.006 | 0.001 | 0.501 |
| 2 | 104214 | 1 | 1 | 0.895 | 0.895 | 0.179 | no | 1 | 0.344 | 0.02 | 0.02 | 0.52 |
| 3 | 8102 | 1.5 | 1 | 0.849 | 0.849 | 0.169 | no | 1 | 0.335 | 0.005 | 0.005 | 0.505 |
| 4 | 71681 | 2 | 1 | 0.833 | 0.833 | 0.166 | yes | 2 | 0.333 | 0 | 0 | 0.5 |
| 5 | 19849 | 2.5 | 1 | 0.854 | 0.854 | 0.17 | no | 2 | 0.333 | 0.013 | 0.013 | 0.513 |
| 6 | 19849 | 2.5 | 1 | 0.854 | 0.854 | 0.17 | no | 2 | 0.333 | 0.013 | 0.013 | 0.513 |
| 7 | 104217 | 3 | 1 | 0.883 | 0.883 | 0.176 | no | 2 | 0.332 | 0.03 | 0.03 | 0.53 |
| 8 | 96100 | 3.5 | 1 | 0.842 | 0.842 | 0.168 | no | 2 | 0.325 | 0.017 | 0.017 | 0.517 |
| 9 | 88601 | 4 | 1 | 0.814 | 0.814 | 0.162 | no | 2 | 0.32 | 0.008 | 0.008 | 0.508 |
| 10 | 99461 | 4.6 | 1 | 0.838 | 0.838 | 0.167 | no | 2 | 0.314 | 0.033 | 0.033 | 0.533 |
| … | … | … | … | … | … | … | … | … | … | … | … | … |
| 2567 | 67155 | 1819.2 | 1 | 0.259 | 0.259 | 0.051 | no | 133 | 0.035 | 0.972 | 0.972 | 1.472 |
| 2568 | 12447 | 1819.9 | 3 | 0.115 | 0.501 | 0.025 | no | 133 | 0.036 | 0.193 | 0.193 | 0.693 |
| 2569 | 20215 | 1820.8 | 1 | 0.156 | 0.156 | 0.031 | no | 133 | 0.035 | 0.375 | 0.375 | 0.875 |
| 2570 | 86184 | 1821.5 | 3 | 0.109 | 0.435 | 0.023 | no | 133 | 0.035 | 0.166 | 0.166 | 0.666 |
| 2571 | 114834 | 1822.3 | 1 | 0.147 | 0.147 | 0.029 | no | 133 | 0.035 | 0.337 | 0.337 | 0.837 |
| 2572 | 86815 | 1823 | 2 | 0.126 | 0.294 | 0.026 | no | 133 | 0.035 | 0.245 | 0.245 | 0.745 |
| 2573 | 57841 | 1823.6 | 4 | 0.09 | 0.766 | 0.02 | no | 133 | 0.035 | 0.091 | 0.091 | 0.591 |
| 2574 | 99701 | 1824.5 | 2 | 0.148 | 0.541 | 0.032 | no | 133 | 0.036 | 0.382 | 0.382 | 0.882 |
| 2575 | 19758 | 1825.4 | 1 | 0.147 | 0.147 | 0.029 | no | 133 | 0.035 | 0.329 | 0.329 | 0.829 |
| 2576 | 101966 | 1826 | 3 | 0.109 | 0.613 | 0.024 | no | 133 | 0.035 | 0.193 | 0.193 | 0.693 |

Notes: HIP is the *Hipparcos* number. $t$ is the current time in mission days. $c$ is the completeness of this observation. $C$ is the total completeness so far, including $c$. P is the probability of a discovery by this observation. Z is the merit function of this observation. $\tau_{LSO}$ is the exposure time for a limiting search observation, in days. $\tau$ is the actual exposure time in days. The last column is the total time cost in days includes overhead. The zeros in Line 4 mean "< $10^{-3}$ days."

Table 3. Design reference mission results for major parameters $D$ and $IWA$ near the baseline.

| case | $D$ | $IWA$ | $\lambda/D$ | observations | revisits | fill | stars in | stars obs | $N_1$ | $N_5$ |
|---|---|---|---|---|---|---|---|---|---|---|
| 1 | 128 | 0.04" | 16 | 3620 | 0 | 0 | 16873 | 3620 | 105.2 | 372.1 |
| 2 | 64 | 0.04" | 8 | 3507 | 0 | 0 | 16869 | 3507 | 103.6 | 366.1 |
| 3 | 32 | 0.04" | 4 | 2999 | 0 | 0 | 16866 | 2999 | 95.6 | 323.5 |
| 4 | 64 | 0.04" | 16 | 3600 | 1944 | 0 | 4719 | 1656 | 64.2 | 153.7 |
| 5 | 32 | 0.04" | 8 | 3409 | 1822 | 0 | 4721 | 1587 | 61.4 | 147.5 |
| 6 | 16 | 0.04" | 4 | 2609 | 1338 | 0 | 4721 | 1271 | 54.8 | 128.7 |
| 7 | 64 | 0.08" | 32 | 3632 | 3162 | 0 | 598 | 470 | 23.6 | 39.0 |
| 8 | 32 | 0.08" | 16 | 3564 | 3103 | 0 | 594 | 461 | 23.3 | 38.3 |
| 9 | 16 | 0.08" | 8 | 3227 | 2785 | 0 | 597 | 442 | 22.8 | 38.2 |
| 10 | 8 | 0.08" | 4 | 2041 | 1690 | 0 | 591 | 351 | 19.9 | 35.1 |
| 11 | 4 | 0.08" | 2 | 702 | 493 | 0 | 590 | 209 | 12.0 | 23.3 |
| 12 | 64 | 0.16" | 64 | 1185 | 1108 | 1233 | 77 | 77 | 3.4 | 4.9 |
| 13 | 32 | 0.16" | 32 | 1200 | 1122 | 1222 | 78 | 78 | 3.4 | 4.8 |
| 14 | 16 | 0.16" | 16 | 1215 | 1136 | 1197 | 79 | 79 | 3.5 | 5.1 |
| 15 | 8 | 0.16" | 8 | 1192 | 1114 | 1101 | 78 | 78 | 3.5 | 5.0 |
| 16 | 4 | 0.16" | 4 | 1145 | 1067 | 51 | 78 | 78 | 3.2 | 4.6 |
| 17 | 32 | 0.31" | 64 | 132 | 124 | 1766 | 8 | 8 | 0.4 | 0.6 |
| 18 | 16 | 0.31" | 32 | 128 | 120 | 1768 | 8 | 8 | 0.4 | 0.6 |
| 19 | 8 | 0.31" | 16 | 123 | 115 | 1769 | 8 | 8 | 0.5 | 0.6 |
| 20 | 4 | 0.31" | 8 | 119 | 111 | 1758.7 | 8 | 8 | 0.4 | 0.6 |

Notes: Each line is the average of 240 design reference missions with the same parameters. Line 6: baseline case. Lines 9–11: cases neighboring the baseline. Other lines: other cases of $D$ and $IWA$ selected to explore the parametrics in the vicinity of the baseline. "Stars in" means all stars in the input catalog with nonzero completeness. "Stars obs" means all unique stars that are observed at least once in a typical design reference mission.

Table 4. Results for offsets of occurrence probability η

| η | $N_1$ | $N_5$ | observations | stars | revisits |
|---|---|---|---|---|---|
| 0.8 | 212 | 517 | 2207 | 1071 | 1136 |
| 0.4 | 106 | 254 | 2467 | 1212 | 1255 |
| 0.2 | 54 | 129 | 2612 | 1272 | 1340 |
| 0.1 | 29 | 67 | 2685 | 1297 | 1388 |
| 0.05 | 14 | 31 | 2736 | 1309 | 1427 |

Table 5. Results for offsets of zodiacal light Z

| Z | $N_1$ | $N_5$ | observations | stars | revisits |
|---|---|---|---|---|---|
| 0 | 60 | 144 | 2864 | 1420 | 1444 |
| 3.5 | 54 | 139 | 2627 | 1273 | 1354 |
| 7 | 56 | 131 | 2703 | 1323 | 1380 |
| 14 | 54 | 125 | 2571 | 1263 | 1308 |
| 25 | 52 | 121 | 2233 | 1076 | 1157 |
| 50 | 47 | 106 | 1952 | 947 | 1005 |
| 100 | 43 | 95 | 1665 | 798 | 867 |
| 200 | 36 | 79 | 1373 | 675 | 698 |

Table 6. Results for offsets of end-to-end efficiency h

| h | $N_1$ | $N_5$ | observations | stars | revisits |
|---|---|---|---|---|---|
| 0.8 | 59 | 144 | 3289 | 1531 | 1758 |
| 0.4 | 58 | 138 | 3015 | 1426 | 1589 |
| 0.2 | 55 | 129 | 2609 | 1271 | 1338 |
| 0.1 | 46 | 108 | 2080 | 1057 | 1023 |
| 0.05 | 39 | 91 | 1484 | 810 | 674 |

Table 7. Results for offsets of scattered light ζ

| z | $N_1$ | $N_5$ | observations | stars | revisits |
|---|---|---|---|---|---|
| $10^{-11}$ | 57 | 133 | 2831 | 1340 | 1491 |
| $10^{-10}$ | 55 | 128 | 2606 | 1274 | 1332 |
| $10^{-9}$ | 41 | 97 | 1604 | 900 | 704 |
| $10^{-8}$ | 19 | 44 | 533 | 366 | 167 |
| $10^{-7}$ | 6 | 12 | 136 | 102 | 34 |

Table 8. Results for offsets of observational overhead $t_{o/h}$

| $t_{o/h}$ | $N_1$ | $N_5$ | observations | stars | revisits |
|---|---|---|---|---|---|
| 0 | 114 | 207 | 8067 | 1890 | 6177 |
| 0.125 | 90 | 177 | 5388 | 1621 | 3767 |
| 0.25 | 72 | 155 | 4003 | 1473 | 2530 |
| 0.5 | 55 | 129 | 2609 | 1271 | 1338 |
| 1 | 38 | 96 | 1538 | 1011 | 527 |
| 2 | 23 | 68 | 841 | 748 | 93 |

Table 9. Results for offsets of the noise floor $\Delta mag_0$

| $\Delta mag_0$ | $N_1$ | $N_5$ | observations | stars | revisits |
|---|---|---|---|---|---|
| 27.5 | 43 | 104 | 1030 | 822 | 208 |
| 27.0 | 54 | 126 | 1505 | 1044 | 461 |
| 26.5 | 57 | 134 | 2069 | 1203 | 866 |
| 26.0 | 55 | 129 | 2609 | 1271 | 1338 |
| 25.5 | 50 | 110 | 3054 | 1129 | 1925 |
| 25.0 | 40 | 86 | 3337 | 917 | 2420 |
| 24.0 | 23 | 43 | 3567 | 489 | 3078 |
| 23.0 | 10 | 15 | 3627 | 203 | 3424 |

Table 10. Results s for offsets of the spectroscopic resolving power $\mathcal{R}_{spc}$

| $\mathcal{R}_{spc}$ | $N_1$ | $N_5$ | observations | stars | revisits |
|---|---|---|---|---|---|
| 17.5 | 56 | 135 | 2772 | 1305 | 1467 |
| 35.0 | 55 | 129 | 2730 | 1298 | 1432 |
| 70.0 | 55 | 129 | 2609 | 1271 | 1338 |
| 140. | 50 | 120 | 2321 | 1190 | 1131 |
| 280. | 42 | 101 | 1666 | 1000 | 666 |

Table 11. Results for offsets of spectroscopic signal-to-noise ratio $S/N_{spc}$

| $S/N_{spc}$ | $N_1$ | $N_5$ | observations | stars | revisits |
|---|---|---|---|---|---|
| 8 | 55 | 129 | 2610 | 1270 | 1340 |
| 20 | 46 | 108 | 1890 | 1066 | 824 |
| 40 | 35 | 80 | 1050 | 771 | 279 |
| 80 | 20 | 43 | 434 | 433 | 1 |
| 160 | 10 | 22 | 205 | 205 | 0 |

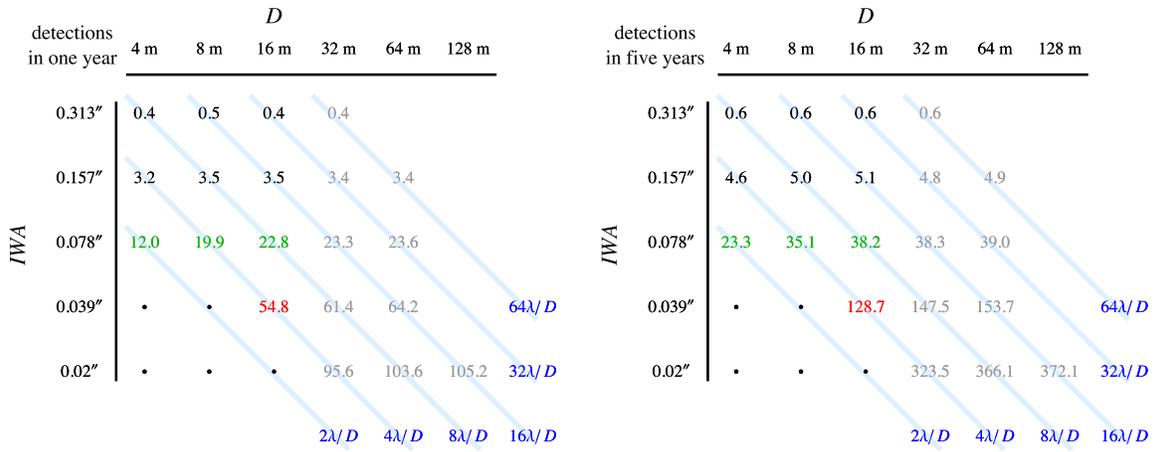

Figure 1. Cases of *D* and *IWA*. Values of *IWA* are shown both in arcsec and units of λ/*D*. Red: baseline case. Green: cases neighboring the baseline. Gray: out of scope (*D* > 16m). Black: underperforming cases ($N_5 \leq 5$).

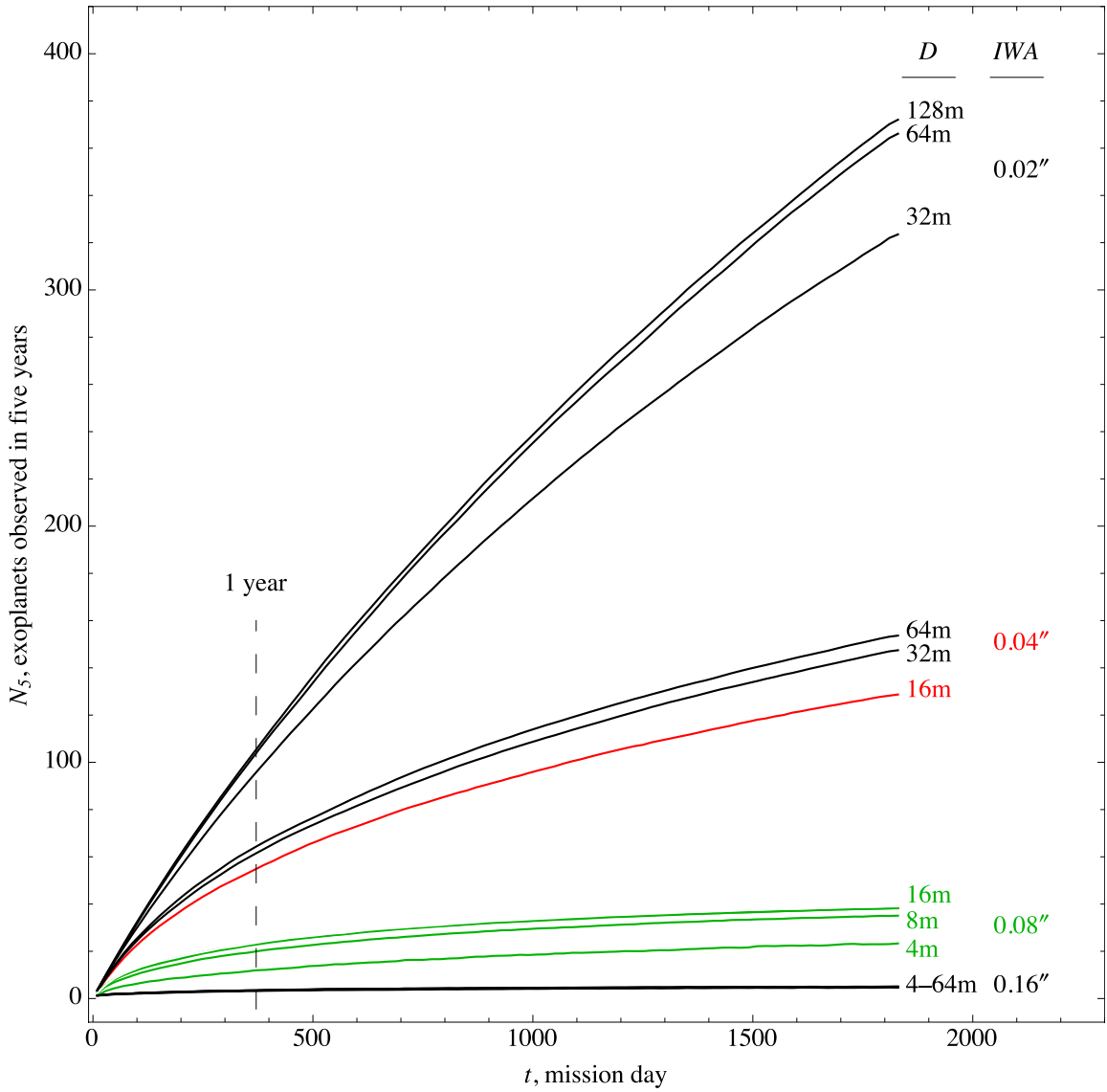

Figure 2. Curves of growth for the capital parameters, *D*, *IWA*, and *t*. Red: baseline case. Green: viable cases that neighbor the baseline. Near the baseline, changing *D* while holding *IWA* constant produces only small changes in the metric *N*.

|     |       | 4m   |      | 8m   |      | 16m   |      | *D* 32m |      | 64m   |      | 128m  |
|-----|-------|------|------|------|------|-------|------|---------|------|-------|------|-------|
|     | 0.08" | 23.3 | 1.51 | 35.1 | 1.09 | 38.2  | 1.00 | 38.3    | 1.02 | 39.0  |      |       |
|     |       |      |      |      |      | 3.37  |      | 3.85    |      | 3.94  |      |       |
| *IWA* | 0.04" | •  |      | •    |      | 128.7 | 1.15 | 147.5   | 1.04 | 153.7 |      |       |
|     |       |      |      |      |      |       |      | 2.19.   |      | 2.38  |      |       |
|     | 0.02" | •    |      | •    |      | •     |      | 323.5   | 1.13 | 366.1 | 1.02 | 372.1 |

Figure 3. Improvements in the metric $N_5$ due to factor-two improvements in *D* and *IWA* near the baseline. Cells with no background: the value of $N_5$ for the values of *D* and *IWA* defining that column and row. Cells with blue background: the improvement factor in $N_5$ going from value of *D* on the left side to the value on the right side. Cells with green background: the improvement factor in $N_5$ going from the value of *IWA* on the upper side to the value on the lower side. The gray numbers are for unreasonably large apertures, which are present only to explore the parametric variations.

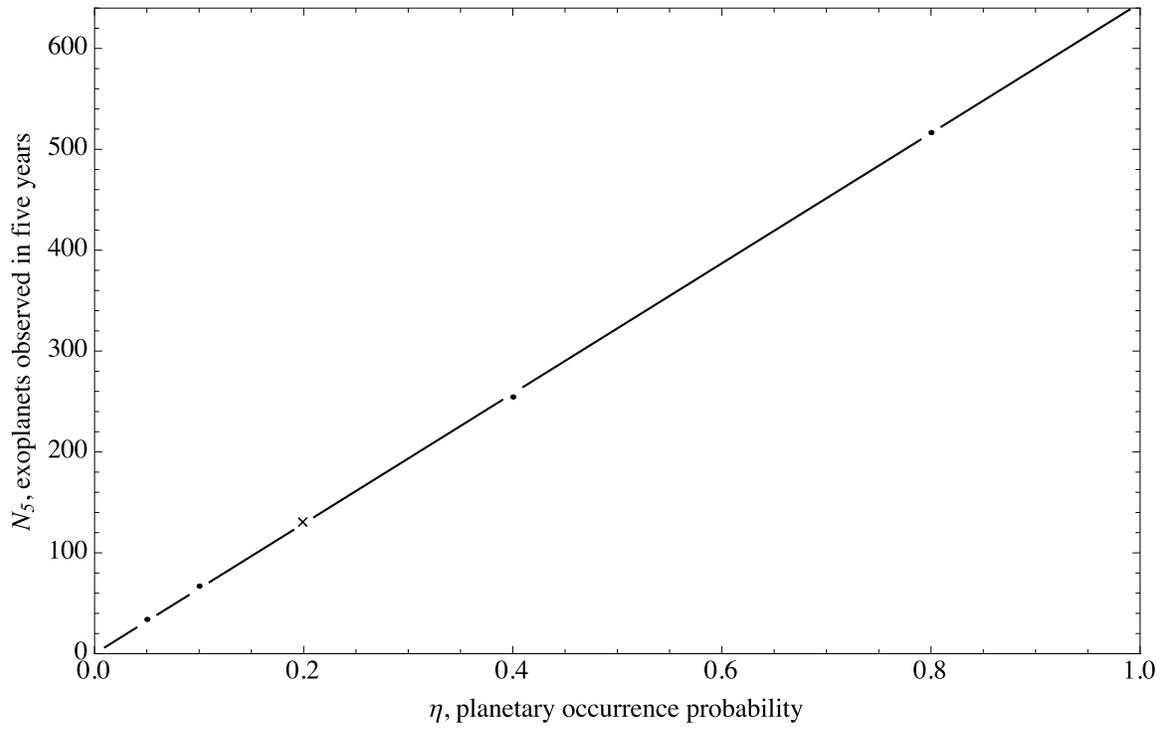

Figure 4. $N_5$ is directly proportional to $\eta$, as expected. Cross: baseline case.

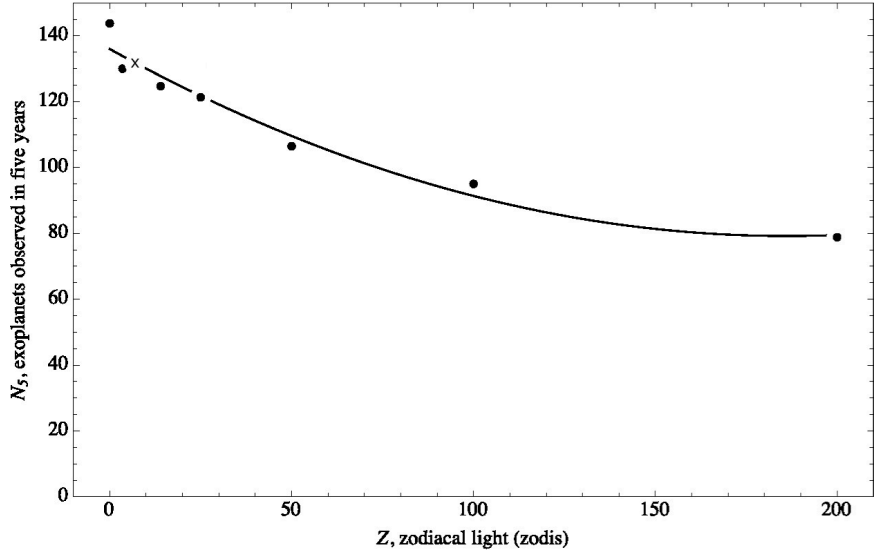

Figure 5. Variation of $N_5$ with zodiacal light $Z$. The effect of $Z$ is weak. Cross: baseline case.

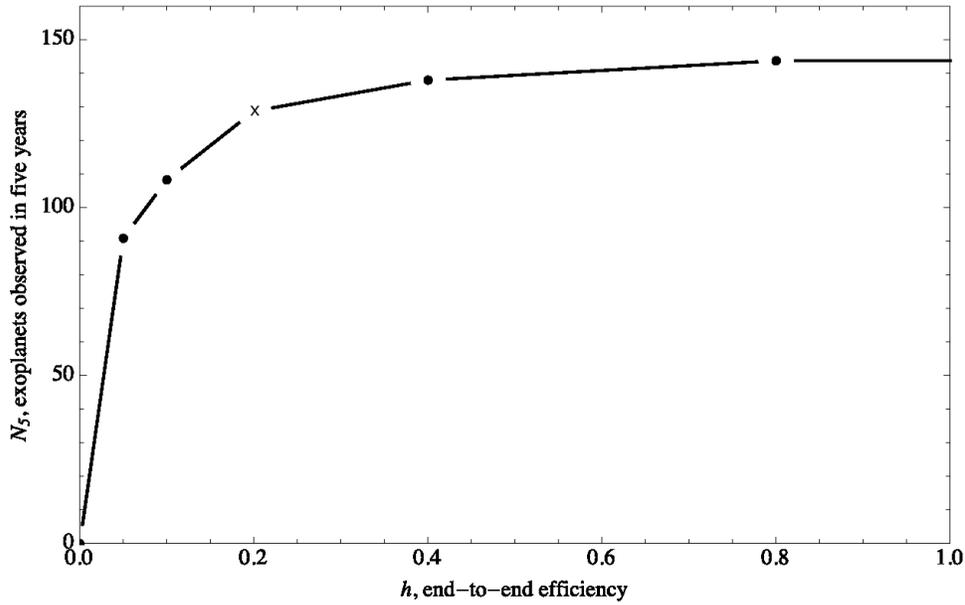

Figure 6. Variation of $N_5$ with end-to-end efficiency $h$. Increasing $h$ above 20% only weakly increases $N$, while decreasing it below 10% drastically reduces $N$. Cross: baseline case.

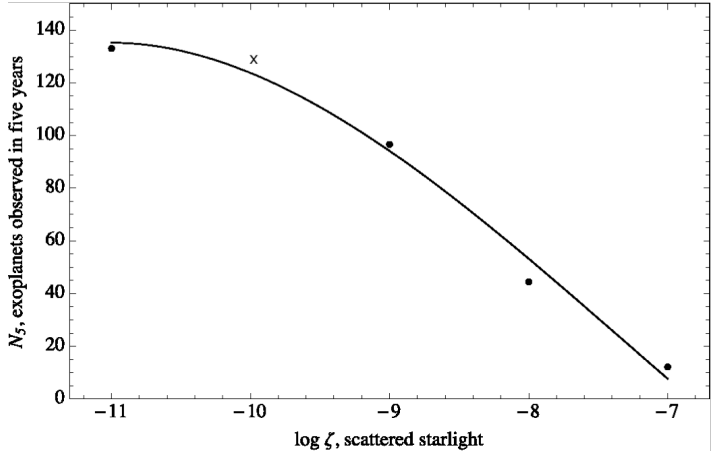

Figure 7. Variation of $N_5$ with scattered starlight, $\zeta$. Increasing $\zeta$ increases exposure time $\tau$, which decreases the total number of observations, which in turn reduces $N_5$. There is little benefit in reducing scattered starlight below $\zeta = 10^{-10}$. Cross: baseline case.

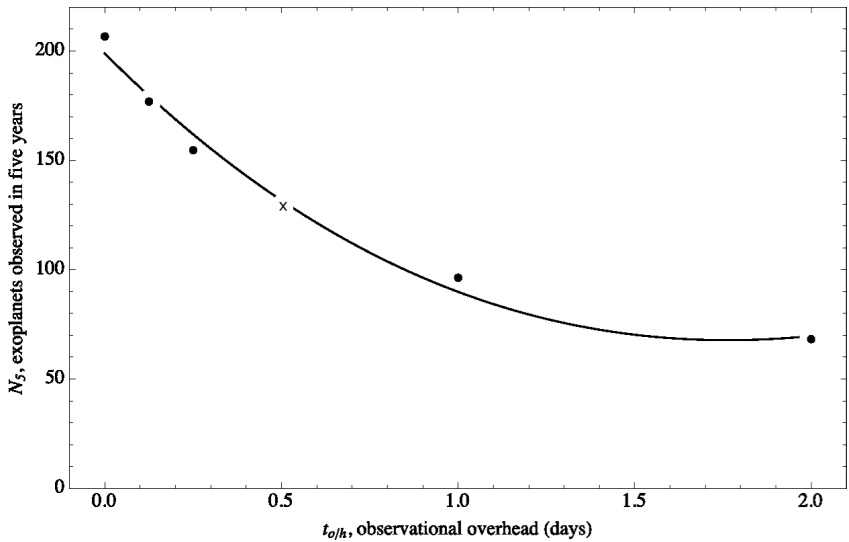

Figure 8. Variation of $N_5$ with observational overhead $t_{o/h}$. Reducing $t_{o/h}$ from 0.5 to 0 days increases $N$ by 50%, while increasing $t_{o/h}$ from 0.5 to 2 days reduces $N$ by 50%. Cross: baseline case.

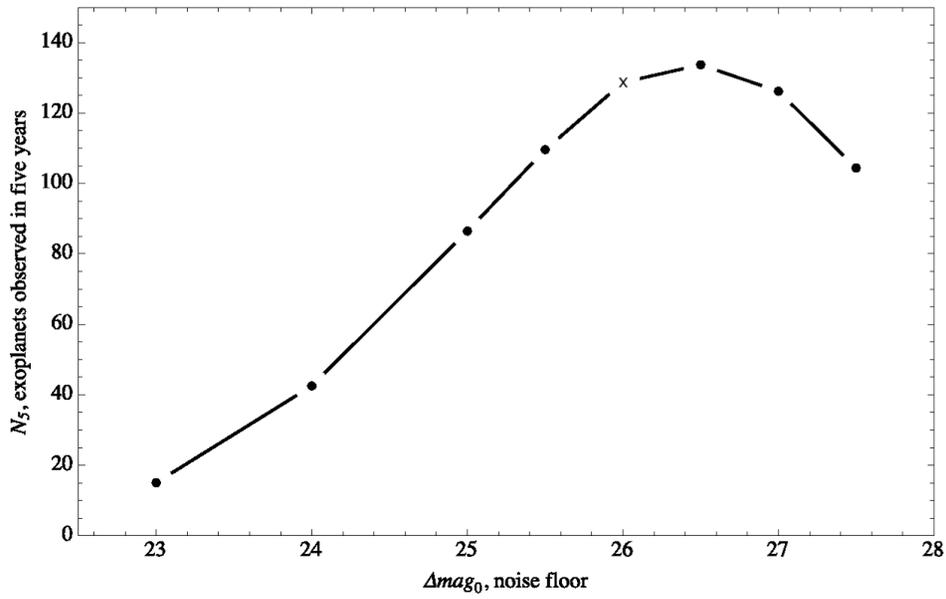

Figure 9. Variation of $N_5$ with noise floor $\Delta mag_0$. The optimal value is $\Delta mag_0 = 26.5$. Cross: baseline case.

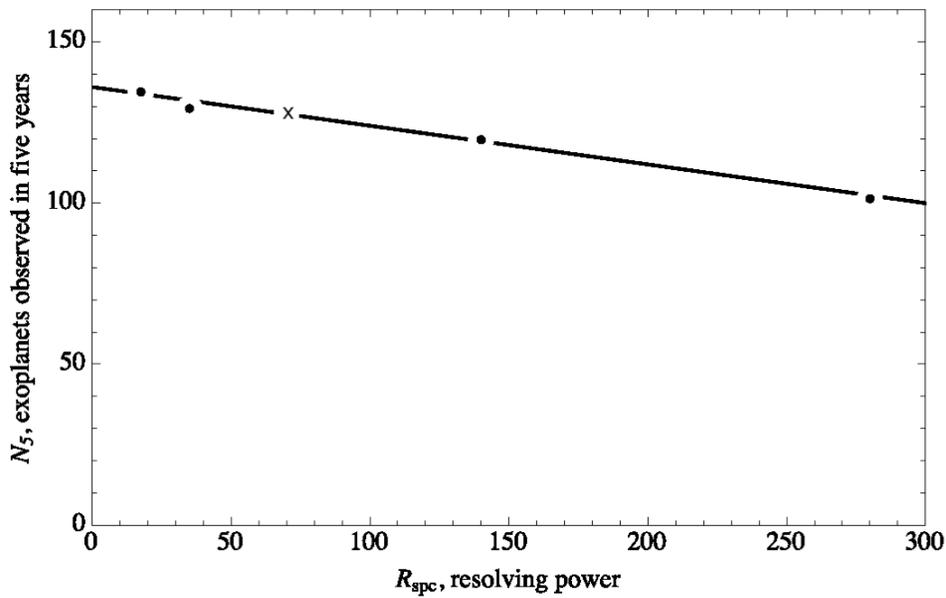

Figure 10. Variation of $N_5$ with spectroscopic resolving power $\mathcal{R}_{spc}$. Cross: baseline case.

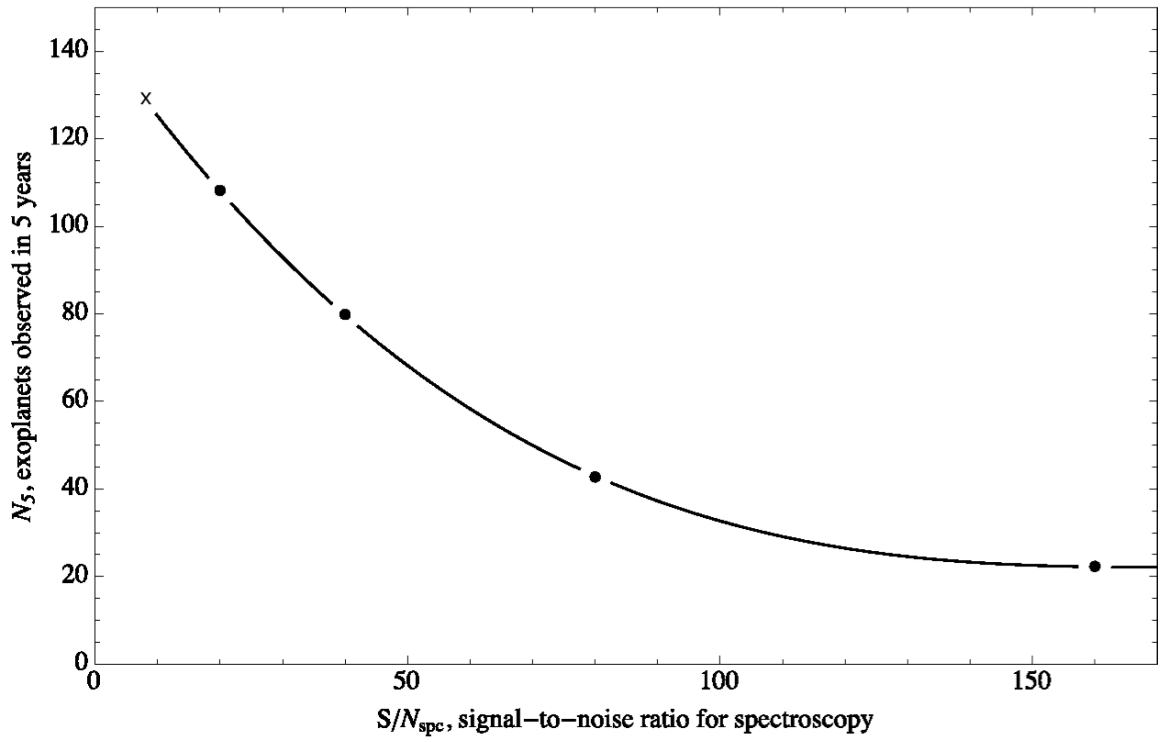

Figure 11. Variation of $N_5$ with spectroscopic signal-to-noise ratio for spectroscopy $S/N_{spc}$. $N_5$ is inversely proportional to $S/N_{spc}$ in this range. Cross: baseline case.